\documentclass[%
 reprint,
 amsmath,amssymb,
 aps,
]{revtex4-2}

\usepackage{graphicx}
\usepackage{dcolumn}
\usepackage{mathrsfs}
\usepackage{bm}
\usepackage[dvipsnames]{xcolor}
\begin{document}

\preprint{APS/123-QED}

\title{Hypersonic acoustic wave control via hyperuniform phononic nanostructures}

\author{Michele Diego$^1$}
\email{diego@iis.u-tokyo.ac.jp}
\author{Jade Hardouin$^{1,2}$}
\author{Gabrielle Mazevet-Schargrod$^{1,3}$}
\author{Matteo Pirro$^1$}
\author{Byunggi Kim$^{1,4}$}
\author{Roman Anufriev$^{5}$}
\author{Masahiro Nomura$^{1}$}
\email{nomura@iis.u-tokyo.ac.jp}

\affiliation{$^1$Institute of Industrial Science, The University of Tokyo, Tokyo 153-8505, Japan}
\affiliation{$^2$École Polytechnique, Palaiseau 91120, France} 
\affiliation{$^3$ESIEE Paris Cité Descartes, Noisy-le-Grand 93160, France}
\affiliation{$^4$Department of Mechanical Engineering, School of Engineering, Institute of Science Tokyo, Tokyo 152-8550, Japan}
\affiliation{$^5$Laboratory for Integrated Micro and Mechatronic Systems, CNRS-IIS IRL 2820, The University of Tokyo, Tokyo 153-8505, Japan}

%
%


\date{\today}

\begin{abstract}
Controlling hypersonic surface acoustic waves is crucial for advanced phononic devices such as high-frequency filters, sensors, and quantum computing components.
While periodic phononic crystals enable precise bandgap engineering, their ability to suppress acoustic waves is limited to specific frequency ranges. 
Here, we experimentally demonstrate the control of surface acoustic waves using a hyperuniform arrangement of gold nanopillars on a lithium niobate layer. The hyperuniform structure exhibits characteristics of both random and ordered systems, leading to an overall reduction in acoustic transmission and the formation of bandgap-like regions where phonon propagation is strongly suppressed. 
We further demonstrate effective waveguiding by incorporating linear and S-shaped waveguides into the hyperuniform pattern. Both simulations and experiments confirm high transmission through the waveguides at frequencies within the bandgaps, demonstrating the flexibility of hyperuniform structures to support waveguides of complex shapes. These findings provide a novel approach to overcoming the limitations of traditional phononic crystals and advancing acoustic technologies in applications such as mechanical quantum computing and smartphone filters.

\end{abstract}

\maketitle


\section{\label{sec:level1}Introduction}
The manipulation of hypersonic acoustic waves is fundamental for applications in sensors \cite{bonhomme2019love, mandal2022surface}, optomechanics \cite{safavi2014two, mirhosseini2020superconducting, kim2023diamond}, topological phononics \cite{zhang2021topological, nii2023imaging, zhang2024monolithic}, and mechanical quantum computing \cite{satzinger2018quantum, dumur2021quantum, qiao2023splitting}. For example, the latter requires thermal insulation at sub-kelvin temperatures, which involves suppression of acoustic waves at gigahertz frequencies. Currently, well-established systems for nanoscale acoustic wave manipulation are phononic crystals \cite{sledzinska20202d, nomura2022review, florez2022engineering, diego2024tailoring}, structures consisting of an ordered periodic array of artificially fabricated scatterers. At the nanoscale, these scatterers are typically implemented as holes in thin membranes \cite{graczykowski2015phonon, diego2024phonon} or pillars on the surface of substrates \cite{pourabolghasem2014experimental, anufriev2023impact}.
However, phononic crystals typically target a specific frequency window rather than a wide frequency range \cite{maldovan2015phonon}. Moreover, this target window becomes even narrower due to slight imperfections in their practical implementation \cite{pourabolghasem2014physics}. Thus, resistance to perturbations remains both a fundamental question and a practical bottleneck in phononics \cite{still2008simultaneous, achaoui2013local, wagner2016two, babacic2024imperfect}. 

In this context, hyperuniform structures are a unique class of materials formed by scatterers arranged with a distribution that falls between order and disorder \cite{torquato2003local}. This behavior is typically described by the structure factor $S(k)$, which reflects density correlations in the reciprocal $k$-space. 
At short range, the distribution exhibits spatial correlations, indicating a degree of organization among nearby scatterers. This results in peaks in $S(k)$ at $k$-vectors associated with the average distance between scatterers.
Conversely, at long ranges, the pattern displays an increasingly uniform distribution of scatterers per unit area, resulting in the suppression of density fluctuations. This long-range uniformity in the real space causes $S(k)$ to approach zero as the $k-$vector tends to zero \cite{torquato2018hyperuniform}. 
One particularly interesting class of hyperuniform structures requires $S(k)=0$ below a certain $k=K$ threshold \cite{torquato2018multifunctional}. This class is called ``stealthy'' hyperuniform and it is widely used in optics, due to large and robust photonic bandgaps \cite{florescu2009designer, froufe2016role, milovsevic2019hyperuniform, aubry2020experimental, vynck2023light}.
However, in phononics, hyperuniform structures are yet to be explored. To the best of our knowledge, phononic hyperuniform structures have been studied either theoretically \cite{gkantzounis2017hyperuniform} or at sub-MHz frequency range \cite{alhaitz2023experimental}.

In this study, we investigate a hyperuniform acoustic structure composed of nanoscale gold pillars arranged according to a disordered stealthy hyperuniform distribution on a lithium niobate layer. Interdigital transducers (IDTs) are used to excite and detect surface acoustic waves, exploiting the piezoelectric properties of lithium niobate. This technique enables the measurement of hypersonic transmission spectra in the gigahertz range. By comparing wave transmission with and without the gold hyperuniform structure, we show its impact on acoustic wave propagation. Fundamentally, we aim to demonstrate that hyperuniform structures hinder acoustic waves across a broad frequency range, with bandgap-like regions that offer waveguiding capability due to exceptionally strong suppression, thus offering an alternative to phononic crystals.

\begin{figure*}[thb]
\centering
\includegraphics[width=0.975\textwidth]{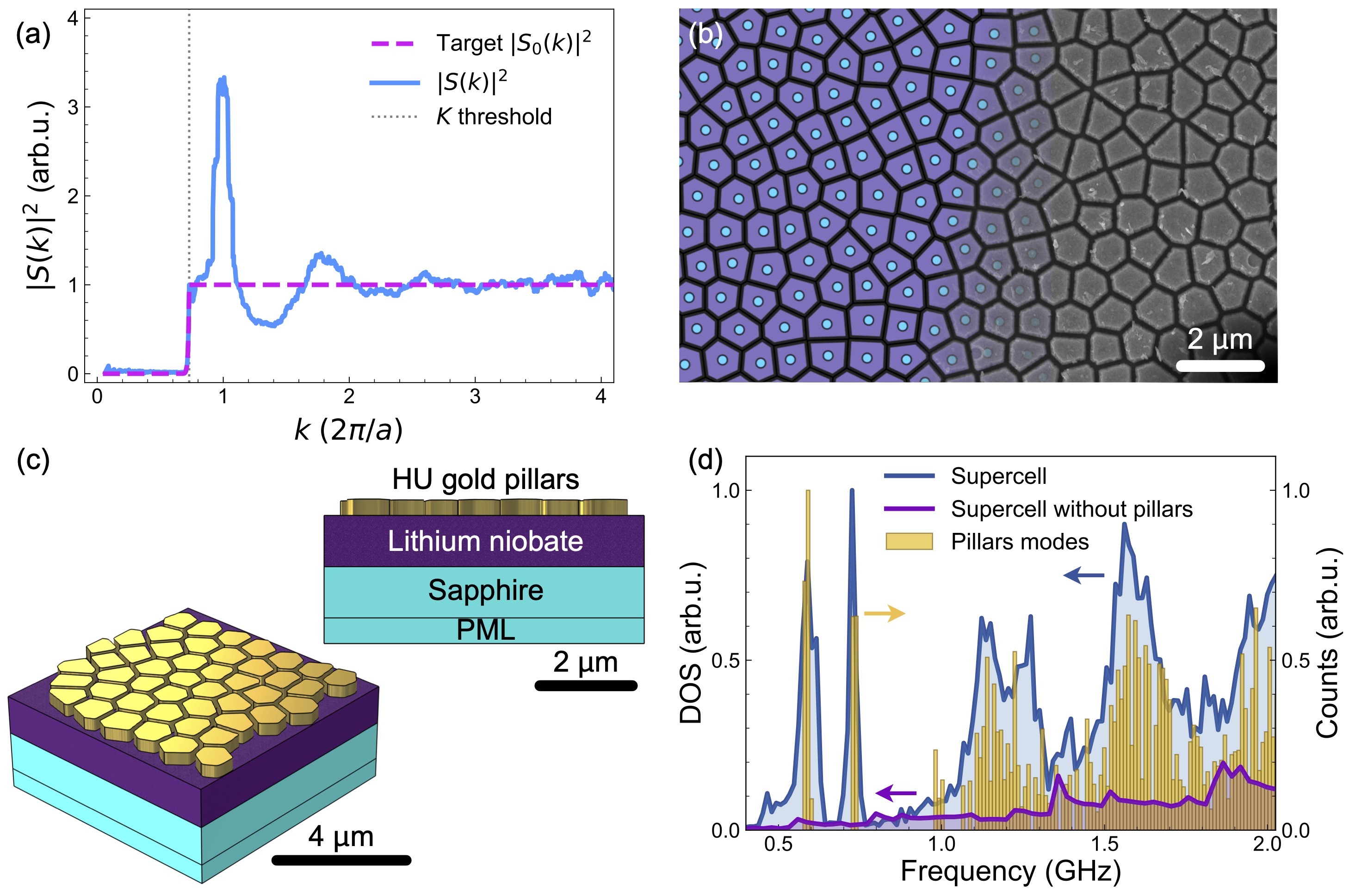}
\caption{Gold hyperuniform nanostructure on lithium niobate. (a) Radially integrated structure factor and target structure factor input into the algorithm to obtain the hyperuniform pattern. In the $x$-axis, $a$ is the average distance between neighbor points. (b) Portion of the hyperuniform distribution (blue dots), its Voronoi cells (purple areas) separated by black walls (left). The pattern gradually transitions into the scanning electron microscope image of the fabricated structure (right), illustrating the connection between the theoretical design and the experimental realization. The structure is formed by gold pillars of 330 nm height deposited according to the hyperuniform pattern on 1-$\mu$m-thick lithium niobate layer upon sapphire. 
(c) A sketch of a portion of the structure, used as an approximate supercell and its lateral cross-section. (d) Density of states of the supercell with and without pillars, and individual pillars modes for comparison.}
\label{fig:1}
\end{figure*}

\section{Results and discussion}
\subsection{Hyperuniform pillars nanostructure}

The hyperuniform structure consists of gold pillars of 330 nm height deposited on a 1-$\mu$m-thick lithium niobate layer atop a sapphire substrate. The pillars are arranged following a stealthy hyperuniform distribution, obtained using an optimization protocol similar to those previously reported in the literature \cite{uche2004constraints}. 
Initially, we provided the optimization algorithm with a disordered distribution of points generated by perturbing a hexagonal lattice. The algorithm computes the structure factor $S(k)$ and evaluates its deviation from a target structure factor $S_0(k)$ designed to enforce a sharp transition from zero to one at $k=K$. It then iteratively adjusts the point distribution to minimize this deviation. The value of $K$ was selected to ensure a high degree of stealthiness in the structure \cite{salvalaglio2024persistent} (see Appendix \ref{apx:hyperuniformity}). Figure \ref{fig:1}a shows the radially integrated structure factor associated with the final hyperuniform distribution, together with the target structure factor used during the optimization process. The peaks in $S(k)$ are expected and derive from short-range correlations between neighboring points.

To assign physical dimensions to the points in the final distribution, we conducted a Voronoi tessellation \cite{torquato2018multifunctional}. This involves partitioning the space into regions, or Voronoi ``cells'', where each cell is centered around a specific point of the hyperuniform distribution and encloses the area that is closer to that point than to any other. 
Finally, to separate the different cells, we introduced walls of uniform width between them, ensuring a consistent distance between the cell borders (see Appendix \ref{apx:hyperuniformity}).

Figure \ref{fig:1}b shows on the left a portion of the final hyperuniform point distribution and Voronoi cells, separated by walls. The image transitions to semi-transparency toward the right, gradually unveiling a scanning electron microscopy image of the actual experimental sample extending to the edge. The sample is obtained by drawing the Voronoi cells pattern using electron beam lithography, followed by the deposition of gold. Finally, a lift-off process is performed to remove excess gold and reveal the hyperuniform distribution of pillars with different shapes. 
The hyperuniformity ensures a high packing density of the pillars, with an average distance between their centers of 800-900 nm, and walls having a width of 80-95 nm.

Given the non-periodic distribution of the pillars and their varying shapes, it is not possible to associate the structure with a specific unit cell. However, by considering a sufficiently large supercell with a high number of pillars, an approximate description of the system as a periodic structure can be achieved. Figure \ref{fig:1}c shows a sketch of such a supercell and its lateral view, employed for finite element method simulations. The sapphire substrate is described by a 1 $\mu$m of sapphire upon an absorbing layer (perfectly matched layer, PML) that approximates an infinite domain.
We used this supercell to calculate the phonon dispersion of the structure and retrieve the density of states (see Appendix \ref{apx:phonondisp}). Figure \ref{fig:1}d shows the density of states of the structure both with and without the pillars. The presence of the pillars significantly increases the density of states, which alters the acoustic properties of the lithium niobate layer. Moreover, we plot a histogram of the individual pillar modes. The first two modes, around 0.6 and 0.75 GHz, correspond to the bending modes and the diameter-breathing mode of the pillars. Since all pillars share the same height and have similar average diameters, these modes are nearly identical across all pillars, leading to narrow peaks. In contrast, higher-frequency modes, associated with more complex deformations, exhibit broad peaks. This is because the variations in pillar shapes cause frequency shifts, resulting in modes with a wide dispersion of frequencies. 
Overall, the individual modes closely resemble the peaks observed in the supercell density of states, indicating the significant influence of the individual pillars and their varying shapes on the acoustic properties of the entire structure.

\subsection{Effect on acoustic waves transmission}

\begin{figure*}[tb]
\centering
\includegraphics[width=0.975\textwidth]{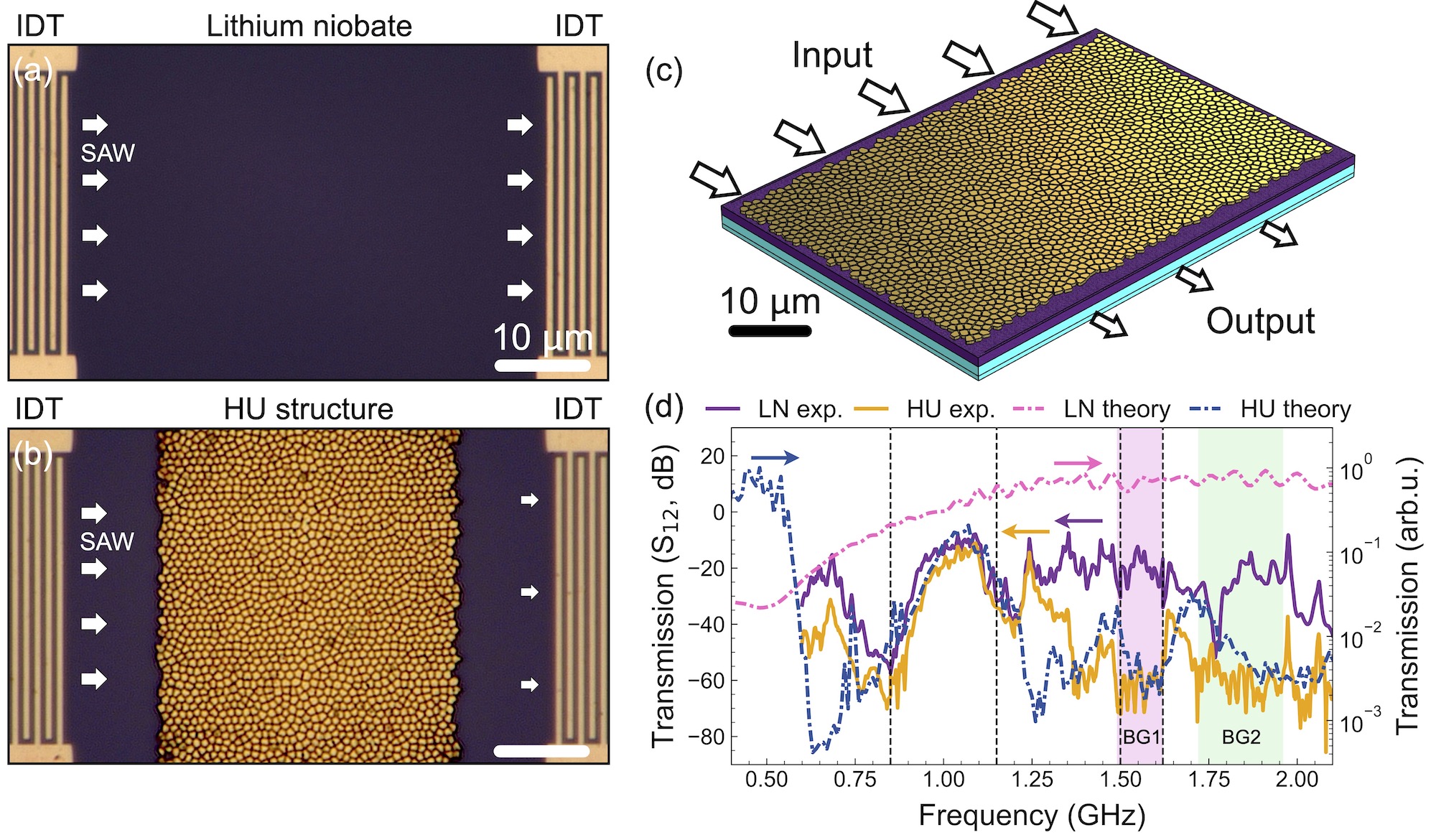}
\caption{Effect of the hyperuniform structure on the transmission of acoustic waves. (a-b) Optical microscope images of the lithium niobate (LN) samples without  and with (b) the gold pillars forming the hyperuniform (HU) structure. At the sides of the structures, IDTs are used to generate and detect surface acoustic waves (SAW).
(c) Same structure as in (b), but in a simulation model to compare experimental and theoretical transmission spectra.
(d) Experimentally measured (left axis) and simulated (right axis) transmission in lithium niobate with and without the gold hyperuniform structure. Vertical dashed lines divide measurements with different IDTs. The experimentally measured bandgap-like regions are highlighted for clarity. } 
\label{fig:2}
\end{figure*}

To evaluate the influence of the hyperuniform structure on acoustic wave propagation in lithium niobate, we measured and simulated transmission spectra of acoustic waves with and without the presence of the hyperuniform pillar structure. For the experimental measurements, we fabricated chirped IDTs designed to excite and detect acoustic waves. Figure \ref{fig:2} shows microscope images of the system with IDTs on the sides, both without (panel a) and with (panel b) the hyperuniform structure.

Simulations were performed via the finite element method. Figure \ref{fig:2}c illustrates the simulation model.
Here, an acoustic excitation with a modulated frequency is applied at the left boundary of the system. For each frequency, the transmission is calculated as the ratio of the elastic energy output on the right side to the input on the left side. This approach mimics the experiment, where one IDT generates the acoustic wave and the other detects the transmission.

Figure \ref{fig:2}d shows the transmission results obtained from both experiments and simulations. The simulation without the hyperuniform structure reveals a gradual increase in transmission within the frequency range of 0.5 - 1.1 GHz, eventually reaching near-complete transmission. The corresponding experimental sample also demonstrates a gradual increase in transmission with frequency, reaching a plateau above 1 GHz. Although a direct quantitative comparison between the simulated and measured spectra is not feasible due to various loss mechanisms present in the real sample, the simulations are generally consistent with the experiments.

The measurements are obtained using five distinct IDTs, each designed to excite a specific frequency range. In Figure \ref{fig:2}d, the measurements from these IDTs are combined, with each dataset cut and joined to the next. Vertical lines indicate the transition points between measurements from different IDTs. The criterion for determining these transition points is to select, for each frequency range, the IDT that exhibits the strongest excitation (see Appendix \ref{apx:IDTmeasurements}). Thus, measurements without the hyperuniform structure are used as a reference to reveal the frequency ranges at which the IDTs provide strong excitation. Dips in the transmission spectrum, such as those observed around 0.8 or 1.2 GHz, are attributed to weak excitation by the IDTs rather than being intrinsic properties of the system. Crucially, the 1.25 - 2.1 GHz range, in which the key physics of this study occurs, is excited evenly.

For the system with the hyperuniform structure, both simulations and experimental results reveal a significant modification of the transmission.  
As a general trend, the transmission displays a series of peaks and dips across all frequencies, with the peaks showing progressively lower values as the frequency increases. 
Certain features, such as the experimental peak near 1.25 GHz, deviate between the experimental and simulated results, likely due to imperfections in pillar fabrication and adhesion to the substrate. Nevertheless, the main characteristics are reproduced, highlighting the role of the hyperuniform structure in effectively engineering acoustic wave propagation within the system.

Above 1.35 GHz, both theory and experiment reveal a broad and significant reduction in transmission, with two distinct regions of particularly low values.
Between 1.5 and 1.65 GHz, both theoretical simulations and experimental measurements show near-total transmission suppression, marked by distinct peaks on either side of this range. Similarly, measurements between 1.7 and 1.95 GHz identify another region of very low transmission, while theoretical predictions suggest comparable behavior at slightly higher frequencies (1.85–2.1 GHz).

The pronounced transmission suppression in these two frequency ranges is interpreted as effective acoustic bandgap-like regions in the system, labeled BG1 and BG2.

\subsection{Waveguides}

\begin{figure*}[tb]
\centering
\includegraphics[width=0.975\textwidth]{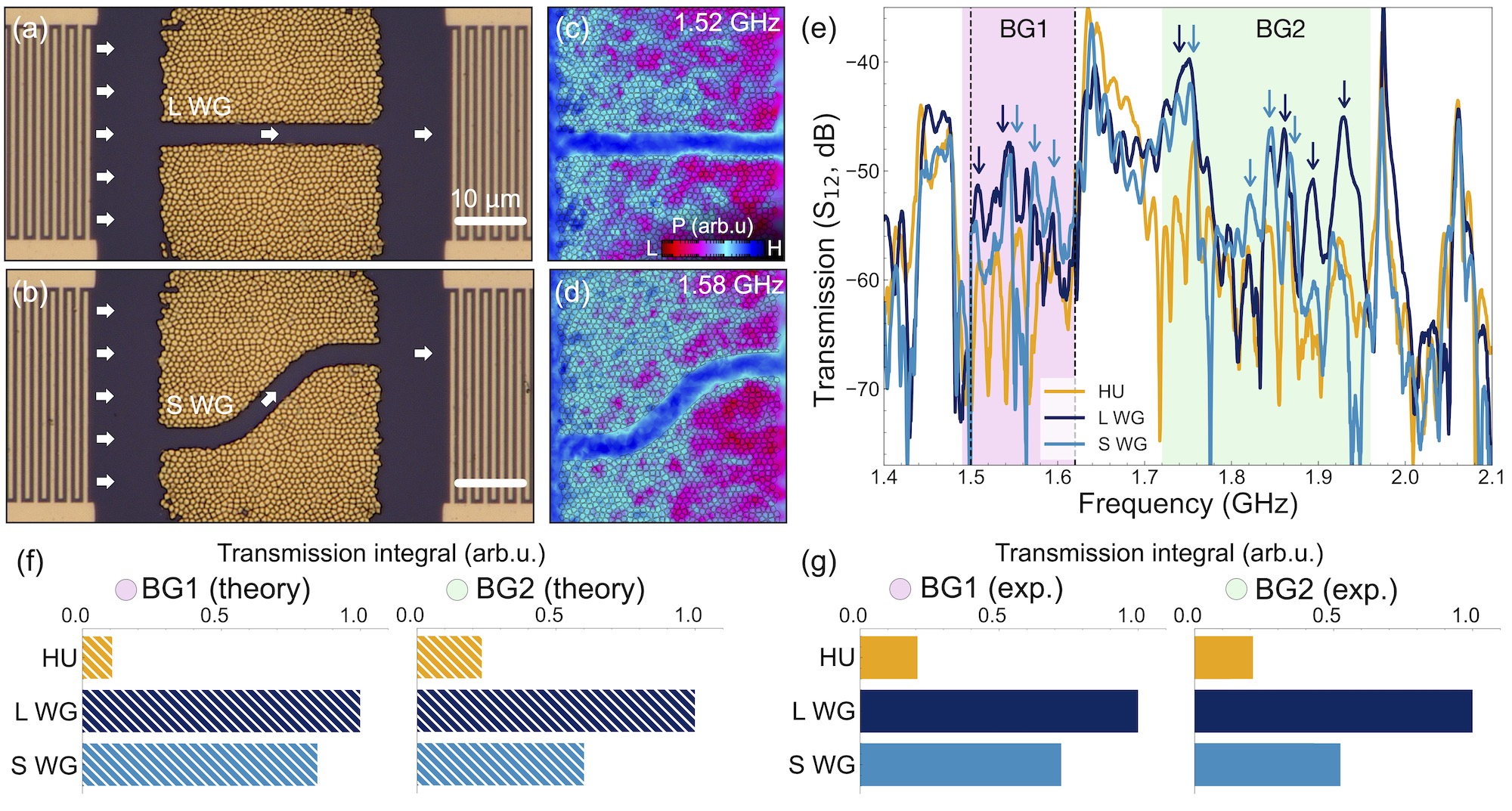}
\caption{Transmission through waveguides. (a-b) Optical microscope images of hyperuniform structure with a linear waveguide (L WG) and a S-shaped waveguide (S WG). (c-d) Simulated Poynting vector ($P$) at frequencies within BG1, in the presence of hyperuniform structures with two waveguide shapes. The excitation is provided at the left of the structure, as in Figure \ref{fig:2}c. These colormaps demonstrate that the presence of the waveguides enables modes with higher transmission.
(e) Transmission measurements (S$_{12}$) comparison between the hyperuniform structure without waveguide and with the two waveguides. Arrows indicate peaks at frequencies of high transmission, associated with modes enabled by the waveguides. (f-g) Histograms of the integrated transmission in simulations (f) and experimental measurements (g) within the effective bandgaps for the hyperuniform structure without and with the waveguides.
}
\label{fig:3}
\end{figure*}

The emergence of effective bandgaps associated with the hyperuniform structure suggests the potential for designing acoustic waveguides capable of transmitting waves at frequencies within these bandgaps. 
Figure \ref{fig:3}a shows a microscope image of the hyperuniform structure with a linear waveguide created by removing pillars along a 3-$\mu$m-width strip. 
Additionally, we designed an S-shaped waveguide, shown in Figure \ref{fig:3}b, to test a freeform waveguide shape. Indeed, as shown in optical experiments \cite{man2013isotropic}, hyperuniform patterns enable the design of freeform waveguides that are not achievable with conventional photonic/phononic crystals.

Using finite element method simulations, we calculated the propagation of acoustic waves in these waveguides, following the same approach as in previous transmission simulations.
Figures \ref{fig:3}c and \ref{fig:3}d present colormaps of the acoustic Poynting vector for the two systems when excited at frequencies within BG1. 
The Poynting vector is significantly higher within the waveguides, suggesting their effectiveness in transmitting acoustic waves.

To observe the wave guidance experimentally, we measured the transmission spectra for the systems with waveguides via IDTs. Figure \ref{fig:3}e compares these transmission spectra with that of the hyperuniform structure without waveguides. Outside the bandgap-like regions, the transmission is similar across all structures, with shared peaks observed at approximately 1.45 GHz, 1.63 GHz, 1.98 GHz, and 2.06 GHz. Within the bandgaps, the transmission is higher in the presence of waveguides, exhibiting distinct peaks marked by arrows. These peaks are interpreted as modes supported by the waveguides.

This result is further demonstrated through the integration of the transmission spectra over the bandgap-like ranges.
Figures \ref{fig:3}f and \ref{fig:3}g present the histograms of transmission (normalized to the highest peak) obtained from simulations and experiments, respectively. For each bandgap, we compare the transmission of the hyperuniform structure and the systems with the two waveguides.

Overall, both simulations and experimental measurements agree that waveguides allow the transmission of acoustic waves. Among all configurations, the linear waveguide consistently achieves the highest transmission. Nevertheless, the S-shaped waveguide also achieves high transmission values within both effective bandgaps.
Note that in both the experiment and simulations, the transmission is averaged across the entire hyperuniform width, since experimentally it is not possible to probe acoustic waves only at the waveguide exit.

These results demonstrate that the hyperuniform structure enables the creation of phononic waveguides, including those with freeform shapes like the S-shaped design.

\section{Conclusion}
This work experimentally demonstrates a practical application of hyperuniformity in nanophononics. Our results illustrate the control of hypersonic surface acoustic waves using a hyperuniform distribution of gold pillars on a substrate.

We first investigated the impact of the hyperuniform pattern on acoustic wave transmission. Our findings reveal that the transmission is significantly reduced by the presence of gold pillars, with experimental measurements and simulations showing overall consistency. Notably, the hyperuniform structure hinders the propagation of acoustic waves over a broad gigahertz range, including bandgap-like regions of particularly strong suppression.

We then exploited the effective bandgaps to design linear and S-shaped waveguides by removing pillars from the hyperuniform structure. Both simulations and experiments confirm that these waveguides allow transmission within the bandgaps. Thus, the hyperuniform structure supports freeform waveguides, such as the S-shaped configuration. This feature offers design flexibility, enabling novel device architectures with improved functionality as compared to traditional periodic phononic crystals.

\begin{acknowledgments}
This work was supported by Japan Science and Technology Agency Moonshot R\&D grant (JPMJMS2062) and by the JSPS KAKENHI (Grant Number JP23KF0203).
\end{acknowledgments}

\appendix
\section{Details on the hyperuniform pattern}
\label{apx:hyperuniformity}
The hyperuniform distribution was generated using $N=418$ points. The threshold $k$-vector was chosen to be $K=\sqrt{8 \pi N}$, a choice that ensures a high level of stealthiness \cite{salvalaglio2024persistent}. This is quantified by the parameter $\chi=0.5$, which reflects the degree of correlation in the system \cite{batten2008classical}. 

To optimize the distribution, we minimized the function:
\begin{equation}
    F = \sum_\mathbf{k} |S(\mathbf{R}_N, \mathbf{k}) - S_0(k)|^2,
\end{equation}
where
\begin{equation}
S(\mathbf{R}_N, \mathbf{k}) = \frac{1}{N} \sum_{i,j=1}^N e^{i(\mathbf{r}_i-\mathbf{r}_j)\mathbf{k}}
\end{equation}
is the structure factor for the specific distribution of points defined by the positions $\mathbf{R}_N = \{ \mathbf{r}_1, \mathbf{r}_2, ..., \mathbf{r}_N  \}$ and the target structure factor is given by
\begin{equation}
S_0(k) =
\begin{cases} 
\ K^{-\alpha} k^{\alpha} & \text{for } 0<k<K \\
1 & \text{for } k \geq K,
\end{cases}
\end{equation}
where we set $\alpha=100$, ensuring that $S_0(k)$ undergoes a sharp transition from near zero to one at $K$.
Note that in the main text, we always use the scalar notation for the wavevector ($k=|\mathbf{k}|$) instead of the vectorial notation. This choice is made for simplicity, as only the magnitude is relevant due to the isotropic nature of the hyperuniform structure.
The minimization process was performed in Python environment using \textit{scipy.optimize.minimize} method.

\begin{figure}[ht]
\centering
\includegraphics[width=0.45\textwidth]{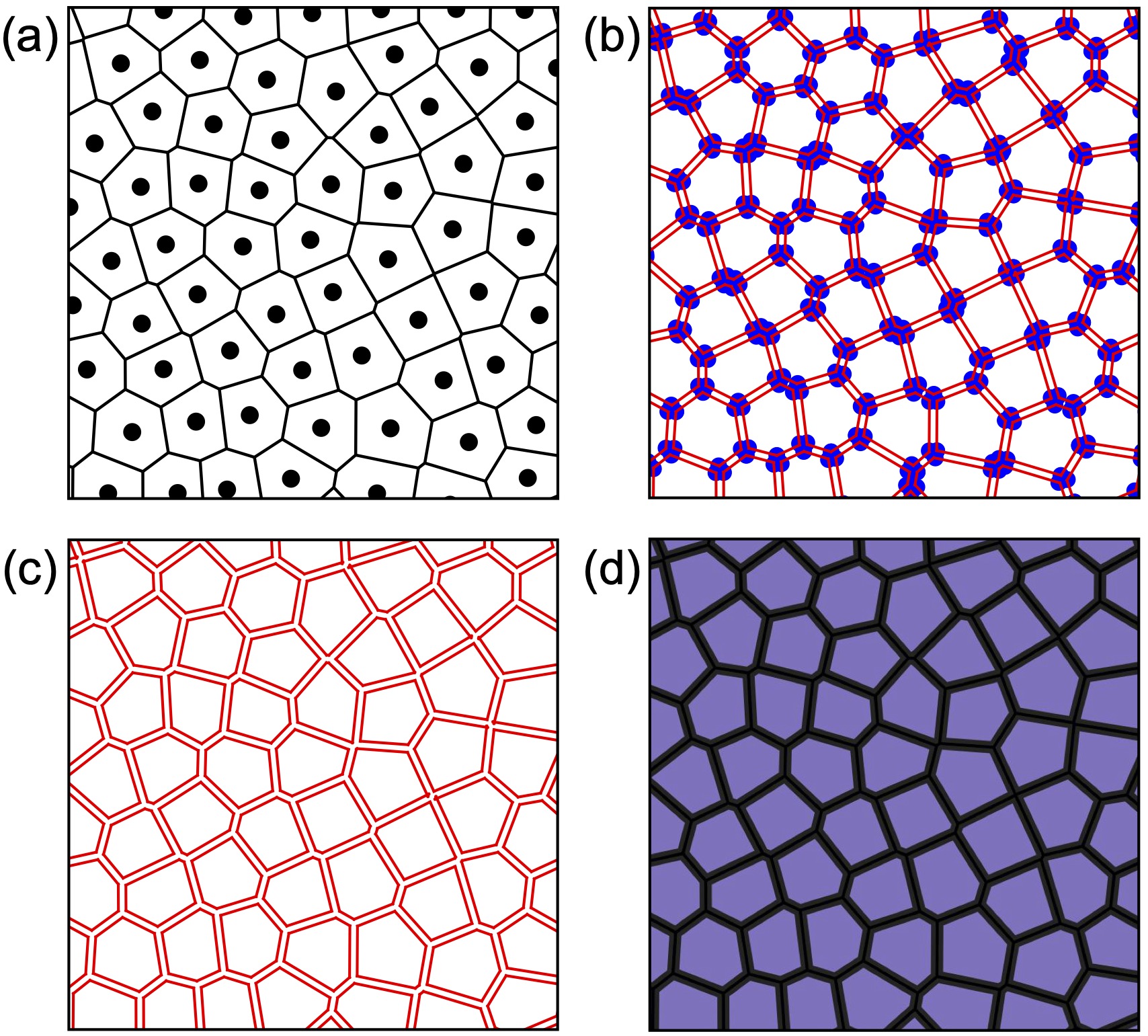}
\caption{Steps for the separation of the Voronoi cells. (a) Point distribution with Voronoi tessellation. (b) Identification of the vertex. (c) Creation of the borders with equal width. (d) Final separated Voronoi cells.} 
\label{fig:4Apx}
\end{figure}

After obtaining the hyperuniform distribution, the Voronoi tessellation was obtained again in Python environment using the \textit{scipy.spatial.Voronoi} method. To create separated cells, we assigned a specified width to the borders of the Voronoi cells. This process involved extensive geometric manipulation using vertex coordinates and edge lists. Figure \ref{fig:4Apx} illustrates some of the steps involved. Ensuring an equal width distance between the cells is crucial for the fabrication feasibility of the structure and for maintaining a uniform average diameter of the resulting cells.

The hyperuniform structure obtained with 418 points yields an experimental sample of approximately 15 $\mu$m$\times$ 15 $\mu$m for an average pillar distance of 800-900 nm. However, to match the scale of the interdigital transducers, the experimental sample dimensions need to be around 30 $\mu$m$\times$ 45 $\mu$m. 
Calculating a hyperuniform structure with that number of points would be computationally prohibitive. Therefore, to achieve the desired dimensions, we assembled a 2$\times$3 matrix of hyperuniform 418-point squares prior to performing the Voronoi tessellation, ensuring smooth transitions between adjacent regions. Consequently, our experimental samples effectively consist of a 2$\times$3 repetition of a large 15 $\mu$m $\times$ 15 $\mu$m supercell containing approximately 418 pillars, with some pillars at the interfaces shared between supercells. 

\section{Details on the finite elements simulations }
\label{apx:phonondisp}
\begin{figure}[ht]
\centering
\includegraphics[width=0.45\textwidth]{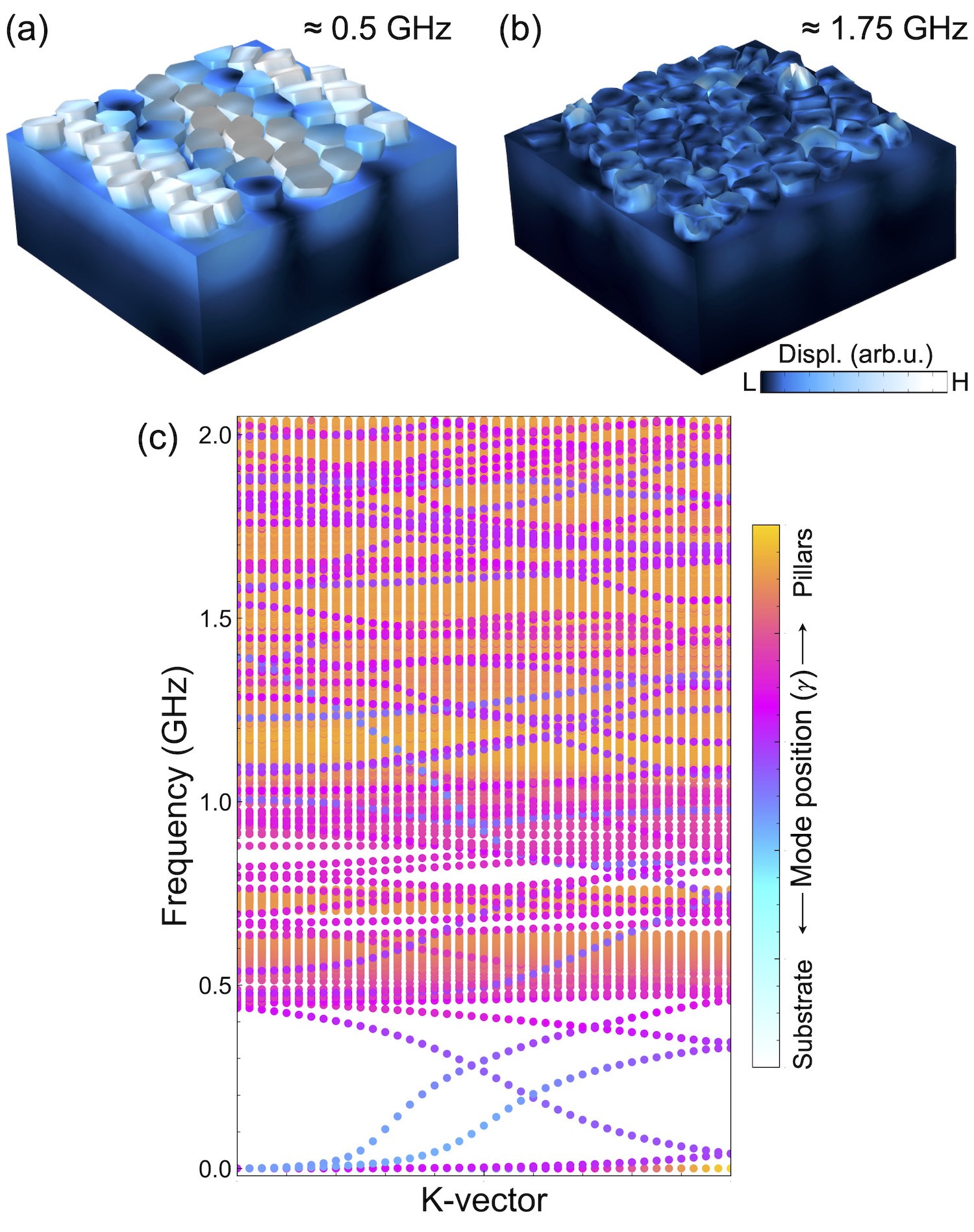}
\caption{(a-b) Displacement for eigenmodes in the supercell at low frequency (a) and high frequency(b). (c) Total phonon dispersion associated with the supercell (Figure \ref{fig:1}c), where each mode is colored according to its vertical position in the cell.} 
\label{fig:5Apx}
\end{figure}

Figure \ref{fig:5Apx} shows the phonon dispersion of the supercell depicted in Figure \ref{fig:1}c. To calculate it, we applied Floquet periodic boundary conditions to the lateral sides of the supercell. We then used finite element method simulations to compute the eigenfrequencies of the system while sweeping $k$-vector in from 0 to $\pi/L_{sc}$, where $L_{sc}= 6.25$ $\mu$m is the length of the supercell side. The phonon dispersion was calculated in only one direction, as the computation is very expensive (taking approximately one week to complete), and hyperuniformity ensures the isotropy of the acoustic properties in different directions. Each point in the phonon dispersion is colored according to
\begin{equation}
    \gamma = \frac{1}{t+h} \frac{\int_V \mathcal{E} \, y \, d\mathbf{r} }{ \int_V \mathcal{E} \, d\mathbf{r}},
\end{equation}
where $t$ is the thickness of the substrate (sapphire + lithium niobate), $h$ is the height of the pillars, $\mathcal{E}$ is the elastic energy density, $y$ is the vertical component of the position vector, and the integrals are performed over the volume $V$ of the supercell.
In this way, $\gamma$ indicates at which height the modes are localized and helps determine whether they correspond to pillar or substrate modes. The phonon dispersion is plotted with substrate modes in front of the pillar modes to enhance the visibility of their features. Peaks (dips) in the theoretical transmission in Figure \ref{fig:2}d approximately correspond to frequency ranges where the modes are more confined within the substrate (pillars).

The density of states shown in Figure \ref{fig:1}d is calculated from the phonon dispersion by grouping the modes based on their frequency, regardless of their $k$-vector. 

To investigate the role of individual pillars in shaping the density of states of the supercell, we performed simulations on a modified supercell configuration where all pillars were removed except for one. For these simulations, the periodic boundary conditions were replaced with absorbing boundaries, since no periodicity is associated with such a system. As a consequence, there is no more sweeping of the $k$-vector.

\begin{figure}[ht]
\centering
\includegraphics[width=0.45\textwidth]{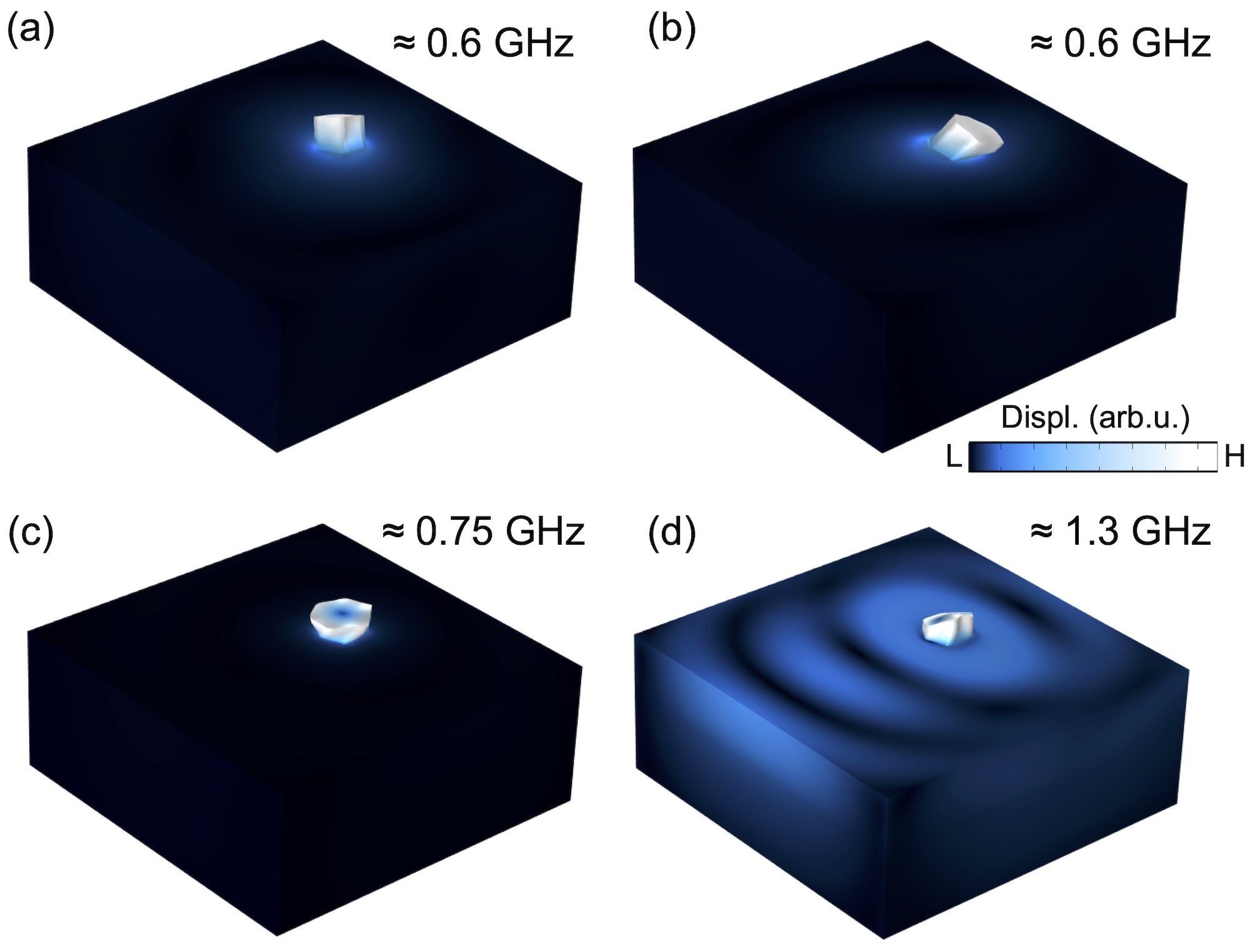}
\caption{Examples of displacement in modes of an individual pillar. (a-c) Low-frequency modes associated with the bending mode of the pillar (a,b) and breathing mode of its diameter (c). (d) High-frequency mode with more complex deformation.} 
\label{fig:6Apx}
\end{figure}

The eigenmodes of the resulting system were computed for several different choices of the remaining single pillar. Figure \ref{fig:6Apx} highlights four modes for a specific pillar configuration. As discussed in the main text, the two lowest modes, at approximately 0.6 GHz, correspond to the bending motion of the pillar (panels a and b). At a slightly higher frequency, around 0.75 GHz, the diameter-breathing mode is observed (panel c). Finally, at approximately 1.3 GHz, a more complex vibrational mode appears.

By performing similar calculations for various choices of the remaining pillar and summing the results, we capture the effect of peak broadening, which originates from the diverse shapes of the individual pillars.

\section{Details on the fabrication of the samples}
\label{apx:nanofabrication}
\begin{figure}[ht]
\centering
\includegraphics[width=0.48\textwidth]{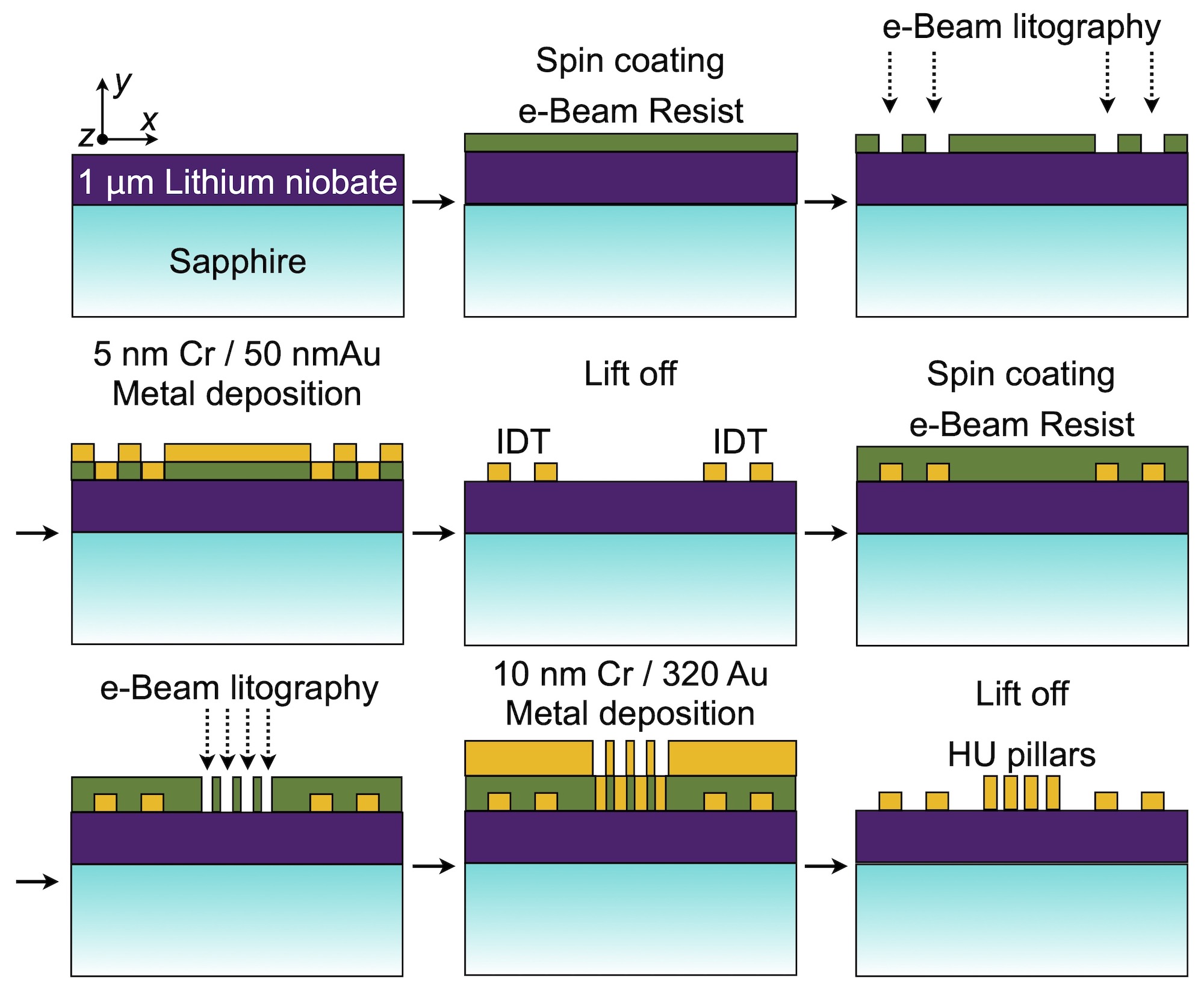}
\caption{Schematics of the fabrication steps.} 
\label{fig:7Apx}
\end{figure}

The wafer used in this work consists of a 1-$\mu$m-thick layer of $y$-cut lithium niobate on a sapphire substrate. The fabrication process involved several steps, summarized in Figure \ref{fig:7Apx}. First, an electron beam resist layer with a thickness of approximately 300 nm was spin-coated onto the sample. Electron beam lithography was then performed to pattern the IDTs, which are oriented to excite acoustic waves propagating along the $x$-axis. Following this, a physical vapor deposition process was used to deposit 5 nm of chromium and 50 nm of gold. A lift-off process was then carried out to obtain the final form of the IDTs.

Five different IDT types were fabricated, each designed to excite distinct frequency ranges. Each IDT consists of 30 fingers on the ground side and 30 fingers on the signal side. The IDTs are chirped, meaning their finger widths were designed to target a central frequency \( f_0 \), with the widths gradually varying to excite a frequency range of approximately \( \pm 15\% \, f_0 \). The distance between the hyperuniform structure and the nearest finger of the IDTs is approximately 10 $\mu$m.

The same sequence of steps was repeated for fabricating the hyperuniform pillars, with slight modifications. In this case, the resist thickness was increased to approximately 550 nm, and the deposition involved 10 nm of chromium followed by 320 nm of gold. In the main text, we approximated the pillars as 330 nm of gold, omitting the thin chromium layer whose only role is the adhesion of gold on the substrate.

\section{Details on interdigital transducers measurements}
\label{apx:IDTmeasurements}

Measurements were performed at room temperature using a vector network analyzer in conjunction with a calibrated microwave probe station.

The parameter $S_{12}$ (or $S_{21}$) describes the transmission coefficient of the device under test, quantifying the amount of signal injected into one port that is transmitted to the other port. 
Conversely, $S_{11}$ (or $S_{22}$) represents the reflection coefficient at each port, indicating how much of the signal is reflected back to the same port.

\begin{figure}[htb]
\centering
\includegraphics[width=0.48\textwidth]{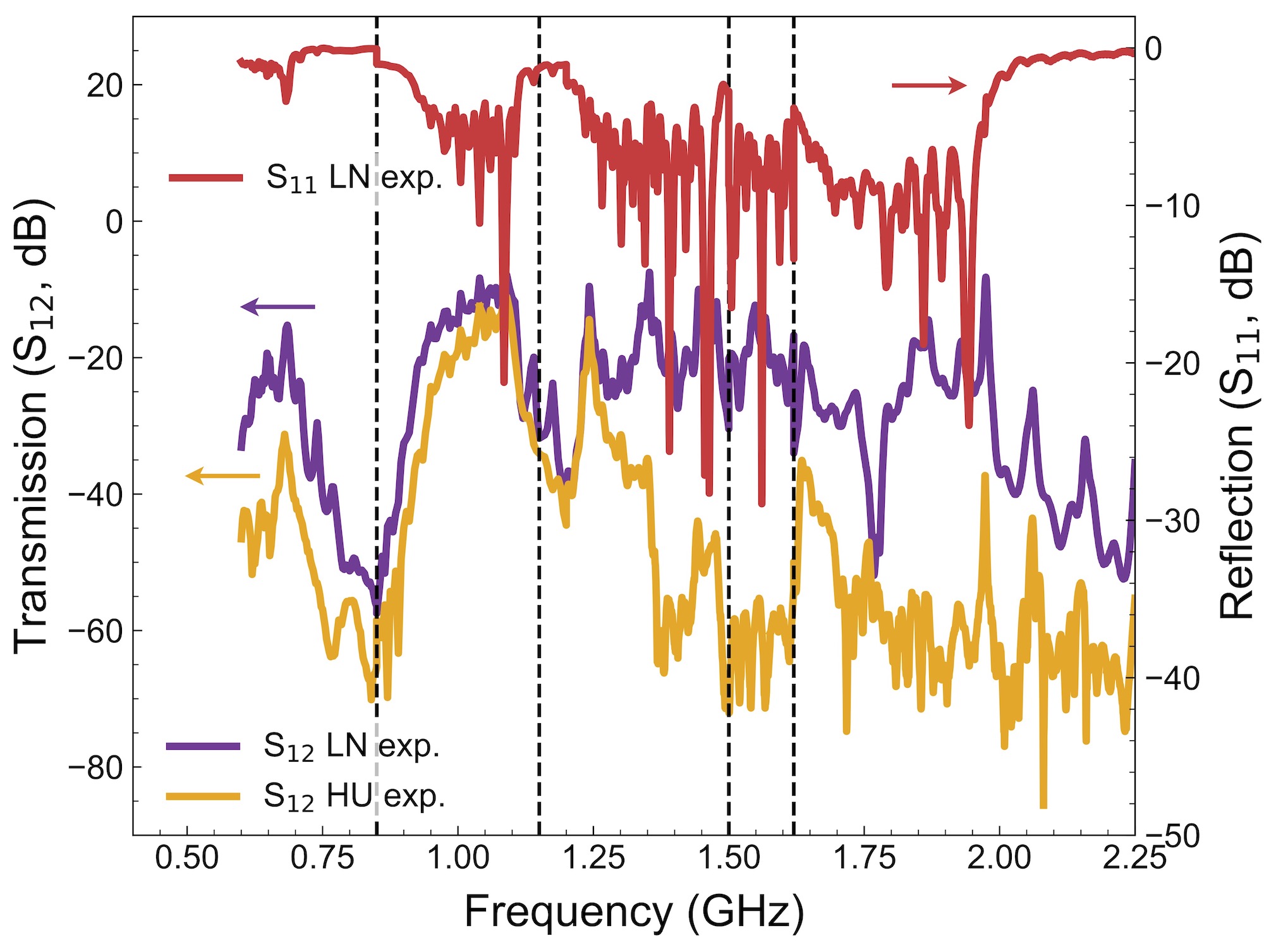}
\caption{IDTs measurements of the transmission (S$_{12}$, left axis) in the presence of the hyperuniform structure, and both transmission (S$_{12}$, left axis) and reflection (S$_{11}$, right axis) without the hyperuniform structure. The reflection shows which ranges of frequencies are properly excited.}
\label{fig:8Apx}
\end{figure}

Figure \ref{fig:8Apx} illustrates the $S_{12}$ parameter both with and without the hyperuniform structure, as well as the $S_{11}$ parameter in the absence of the hyperuniform structure. The $S_{11}$ parameter identifies the frequencies excited by each IDT. The measurements are combined to create a unified, coherent spectrum. The criterion for connecting different measurements is to select, within each frequency range, the data from the IDT that exhibits the highest excitation in the absence of the hyperuniform structure.
In the 0.6–1.2 GHz range, measurements were performed with a step of 2.5 MHz (400 steps per GHz), while in the 1.2–2.25 GHz range, the step size is reduced to 0.5 MHz (2000 steps per GHz).
To calculate the integral of the transmission in the bandgap-like regions in Figure \ref{fig:3}g, we converted the $S_{12}$ data to the transmission in linear scale through $T=10^{S_{12}\text{(dB)}/10}$.

\bibliography{apssamp}

\end{document}